# Computing Traffic Accident High-Risk Locations Using Graph Analytics

Iyke Maduako[1,2], Elijah Ebinne[1], Victus Uzodinma[1], Chukwuma Okolie[3*] and Emmanuel Chiemelu[1]

[1]Department of Geoinformatics and Surveying, Faculty of Environmental Studies, University of Nigeria Enugu Campus, Enugu State, Nigeria
[2]Office of Global Innovation, UNICEF, New York, NY 10017
[3]Department of Surveying and Geoinformatics, Faculty of Engineering, University of Lagos, Lagos State, Nigeria

*Correspondence: iykemadu84@gmail.com

**Abstract**

Analysis of the dynamic relationship between traffic accident events and road network topology based on connectivity and graph analytics offers a new approach to identifying, ranking and profiling traffic accident high risk-locations at different levels of space and time granularities. Previous studies on traffic accident hot spots have mostly adopted spatial statistics and Geographic Information Systems (GIS) where spatial point patterns are discovered based only on spatial dependence with no recognition of the temporal dependence of the events. A limitation arises from the fact that the results are either under or over-estimated because of the temporal aggregation of the events to an absolute time point. Furthermore, the existing methods apart from the Network Kernel Density Estimation (NETKDE), consider traffic accident events as events randomly on a 2-D geographic space. However, traffic accident events are network constrained events that happens majorly on the road network space. Therefore, in this paper, we adopt the connectivity of graph on a network space approach that identifies accident high risk-locations based on space-time-varying connectivity between traffic accident events and the road network geometry. A simple but extensible traffic accident space time-varying graph (STVG) model is developed and implemented for this study. Traffic accident high risk-locations are identified and ranked in space and time using time-dependent degree centrality and PageRank centrality graph metrics respectively through time-incremental graph queries. This study offers urban traffic accident analysts with a new and efficient approach to identify, rank and profile accident-prone areas in space and time at different scales.

**1.      Introduction**

Identification, ranking and profiling of traffic accident high-risk locations have been a major focus of traffic accident analysis by researchers, transport authorities and traffic accident departments [1 – 3]. These agencies are particularly interested in understanding how traffic accidents occur in terms of location (where) and time (when), and the pattern of occurrence over time and space. Several studies have utilized different methods based on spatial point pattern and Geographic Information Systems (GIS) to analyze traffic accident hot spots and hazards. These spatial statistical and GIS methods include, the Moran's I [4], Getis-Ord [5], Kernel Density Estimation (KDE) [6], Full Bayes hierarchical model [7], Quasi-Poisson model [8] and recently Network KDE (NETKDE) [9].

However, the challenge is that these methods are based on spatial statistics where spatial point patterns are analyzed only based on the spatial constraints of the accident locations with no account of the temporal constraints of these events. This leads to either under or over-estimation of the analysis results based on the aggregation of the events to an absolute time point. Therefore, a method that analyses traffic accident hot spots (high risk-locations) from the perspective of connectivity in space and time between road network topology (streets and intersections) and traffic accident events is required to gain more insights on spatio-temporal pattern of accident high risk-locations.

Our approach therefore, differs from the previous approaches in the sense that we identify and rank accident high risk-locations in space and time within a network based on the connectivity between accident events, road network geometry and time using graph data model. Graph data models present fundamental data structure for capturing and analyzing connectivity between events, objects and time. While graphs have been used for decades to analyze interactions in network sciences, its usage to analyze a network-constrained event such as traffic accidents in a transit network remains to be investigated. In this paper, we use a graph data model for identification, ranking and analysis of traffic accident hot spot patterns in space and time at different spatial and temporal granularities.

Our Space and Time-Varying Graph (STVG) model is a network and time constrained model developed in this study to analyze traffic accidents hot spot patterns over the road network entities and across different time granularities. Graph metrics through graph queries are employed to discover traffic accident high-risk locations and analyze their patterns over a period of 6 years (2010-2015) at different time granularity. Degree and PageRank centrality graph metrics are utilized to identify and rank accident-prone segments and intersections. Changes in these metrics over a period of 6 years at various time granularities is used to understand the dynamics in traffic accident pattern in a given location over time. The following analytical queries are computed based on graph metrics from the STVG graph database of traffic accident events: Which intersections and streets have the highest crash index? When and where do most crashes occur? how does spatio-temporal pattern of fatalities differ from spatio-temporal pattern of traffic accidents overall? What is the spatial pattern during weekdays and weekends, morning, afternoon and night times? When and where are the blackspots for crashes involving elderly drivers, teenagers, alcohol or weather related?

The main contribution of this paper is developing a graph data model that can support the computation of graph metrics to identify and rank traffic accident high-risk locations as well as discovery of the space-time varying pattern at different space and time scales. We propose two graph metrics to understand and quantify the connectivity that exists between road network geometry and traffic accident events, facilitating the analysis of road segments and intersections accident activity levels, accident activity ranking and overall accident event pattern over time.

This paper is organized in the following order: a review of past works and methods of traffic accidents analysis is given in section 2. Section 3 describes our proposed Space-Time-Varying Graph (STVG) model for traffic accidents. Graph metrics used to identify and rank the hot spots and temporal pattern analytics are defined in section 4. The practical implementation of the model in the Neo4j graph database is described in section 5. The results of the analysis are presented in section 6 and the conclusion follows in section 7.

## 2. Literature Review

The last two decades have produce vast studies and literature on hot spots and spatial pattern analysis of traffic accidents. Traffic accidents analysis have majorly been studied based on spatial statistical methodologies. This literature review brings as a background, some of the spatial and statistical methods recently used to analyze traffic accidents. Local spatial statistics methods, the Moran's I and Getis-Ord statistics [10, 11] have been popularly used while other studies have adopted methods such as Kernel Density Estimation (KDE) [12, 13], Network KDE (NETKDE) [14, 15], Quasi-Poisson model [8], Full Bayes hierarchical model [16 - 18], Neuro-Fuzzy approach [19, 20]and even Neural Networks [21 – 25]. Moran's I (MI) is a local spatial statistical method that is used in traffic accident analysis to measure the spatial dependence of accident locations and can also be used to examine the density of their spatial pattern, how dispersed or randomly distributed the cluster patterns are. The combination of MI and Getis-Ord statistic have been very effective in furthering the understanding of the processes that lead to spatial dependency [10]. A group of high index Getis-Ord values (z-score and p-values) represent hot spots while the low values are low incident areas. These methods have been used by [1, 26 – 31] for spatial pattern, hot spot detection and severity index of traffic accidents. KDE is one of the most used methods in traffic accident density estimation to examine first order attributes of spatial point dispersion/distribution patterns [32]. Its use has been showcased in the determination of traffic hot spots for example, in measuring spatial concentration and pattern of road accidents [33], wildlife and vehicle accident spatial analysis [34], traffic accident spatial cluster (hot spots) analysis [26], determination of pedestrian crash areas [35], detection of accident hot spots with combined KDE and Poisson function [26], cyclist's traffic hazard density estimation [36], classification of road accidents [37], traffic accident spatial pattern analysis with combined KDE, nearest neighbor distance and K function [38]. While KDE has long been used in traffic accident analysis some authors suggest that it has some limitation based on the premise that KDE's analysis is always constrained to planar 2-D Euclidean space not network space, therefore KDE's density estimation is based on Euclidean distance instead network distance. This has given rise to the development of Network-based KDE (NETKDE) by Xie & Yan [39] to estimate traffic accident density over the network space. NETKDE has been utilized in traffic accident spatial density estimation over the network space [14, 40 – 43]. The introduction of NETKDE showed some useful effort to examine spatial point pattern over the road network geometry, however, traffic accident analysis based on this model focuses on 1-D network space (spatial dimension) with no account of time (temporal dimension).

In view of this background, we have introduced connectivity-based spatial analysis based on space- time graph model that gives us the ability to analysis traffic accident patterns as a function of the interactions between road network geometry, traffic accident events and time. To our knowledge, this is the first work to address traffic accident high-risk location identification, ranking and pattern dynamics over network space from the perspective of graph theory and analytics.

## 3.     Our Proposed Graph Models

There are conceptual and implementation challenges in representing and analyzing the dynamics in traffic accident pattern and high risk-locations based on the network connectivity. The conceptual and logical graph model of these events and the road network topology must assume a flexible structure that is defined in terms of relationships among the entities in both spaces (geographic and event spaces) of the entire network as a function of time. The modelling process thereby involves the logical representation of the interplay between the different perspectives of a traffic accident event network onto a single graph model. At the implementation level, the way in which the traffic accident data is stored, adapting the conceptual graph model in a graph database has clearly an impact on the analytical tasks needed for computing the analysis metrics. Identifying, ranking and analyzing patterns present in a traffic accident event network over time and space are fundamental analytical tasks needed to be supported by a graph database.

### 3.1     Space-time concept of the Model

Conceptually, the Space-Time-Varying Graph (STVG) model of the traffic accident events is a directed property graph model with two major key aspects and two network spaces. The aspects are the network elements (nodes and edges) and the time-tree (time instants). The network elements are spatio-temporal entities on two logical spaces (the geographic and event spaces) of the network. The spatial component of the graph model is a property in the nodes and edges of the network. Therefore, the STVG model can be defined and visualized as a graph, G made up of (N, E, T): where N represents the set of nodes (network vertices), E represents the set of all edges (relationships) and T represents the set of all time instants (from the time-tree).

The set of all nodes in the graph, G is represented as N(G), the set of all edges is represented as E(G) and the set of all time instants is represented as T(G). N(G) is made up of spatial nodes $n_s$ and spatio-temporal nodes $n_{st}$ instants and non-spatio-temporal nodes n. E(G) is consisting of spatio-temporal edges $e_{st}$ and non-spatio-temporal edges e. A spatial node in the graph is a node with spatial properties (such as spatial coordinates) among other properties that it has but with no temporal property, these are the street and intersection nodes (see Fig. 1). Non-spatio-temporal nodes are nodes with no spatial and temporal properties, these are the accident influencing factor nodes. While the spatio-temporal nodes are nodes with spatial and temporal properties in the graph model, that is the traffic crash node. Spatio-temporal edges with the space-time relation "NEXT" in the geographic space and in the event space. The edges can be weighted spatially by distance and temporally by time. and Stops nodes. A spatial edge $e_s$ ∈ E(G) is an ordered triple, that is, $e_s$ = (u, $v$, $w_s$ ), where u, v ∈ N(G) are the source and target nodes, respectively and $w_s$ is the spatial weight of the edge (distance).

A temporal edge $e_t$ ∈ E(G) is an ordered quintuple, $e_t$ = (u, $t_a$ , v, $t_b$, $w_t$ ), where u, v ∈ N(G) are the source and target nodes, respectively, $t_a, t_b$ ∈ T(G) are the source and target time instants, respectively and $w_t$ is the temporal weight of the edge (time). While the spatio-temporal edge $e_{st}$ ∈ E(G) is an ordered sextuple $e_{st}$ = (u, $t_a$ , v, $t_b,$ $w_s, w_t$) where u, v ∈ N(G) are the source and target nodes, respectively, $t_a, t_b$ ∈ T(G) are the source and target time instants, respectively, $w_s$ is the spatial weight of the edge (distance) and $w_t$ is the temporal weight of the edge (time). Non-spatio-temporal edges are edges with no space-time relation and properties such as the "LOCATED_AT", the "CAUSEDBY" and "HAPPENS_AT" edges. This is an edge, $e_s$ ∈ E(G), that is, $e_s$ = (u, $v$), where u, v ∈ N(G) are the source and target nodes, respectively.

### 3.2 Logical graph representation of the Model

We propose a graph model as a data structure that consists of nodes (also known as vertices) and edges (also known as links). The reasons for presenting every element in our graph model as a node are, (i) to accommodate the space-time dynamics of every node in the network and (ii) to enable the computation of relevant graph metrics to determine high risk-locations and accident activity levels which are majorly nodal metrics. The nodes consist of entities that serve as the basis of the representation and edges act as relationships between these entities. A node can be an entity can be a real-world feature that exists in a traffic accident event network such as the geographic features (streets and intersections). A node can also represent a purely conceptual and abstract entities such as the traffic crash and the influencing factors. In our graph model, the edges are used to represent the directed relationships between the entities in the network.

## 3.3	Geographical and Event Spaces

In this research work, we propose the design of a graph model for the connectivity interplay between traffic accident events and geographic spaces on the road network topology where nodes and edges can be combined to form complex retrieval criteria based on graph metrics such as centrality metrics. Our approach proposes superimposing two perspectives of network using a single graph model. The first perspective is the relative view of the road network using the Geographical Space. In our case, the Geographical Space is a topological space based on the existence of geographical places and their neighborhood relationships such as left and right or being connected with. This represents the sequence of places (locations) on the road network, the traffic crashes are located at, sequentially and temporally as they occur. This sequence of occurrence is captured in both spaces of the model by the "NEXT" space-time relationship (edge). The second perspective is the relative view of the traffic crashes on the event space. The Event Space is an abstract space that records the sequence of traffic crashes as they occurred. These traffic crashes are recorded in the event space of the graph database sequentially and are connected to each other by the "NEXT" space-time relation.

It is crucial to note that the geographical and event spaces have complementary characteristics which are evident in their many similarities and important differences. For example, each node in the event space has an equivalent node in the geographical space. This connection between the two spaces is represented by the "LOCATED_AT" relationship. In terms of similarities, they are both non-metric spaces representing the connectivity between geographical places and abstract traffic crash entities. Connectivity is a central characteristic that is being used as the basis of the representation of nodes and edges in both Geographical and Event Spaces. Irrespective of the space that is represented in the graph model, a temporal relationship (HAPPENS_AT) is used to connect all nodes to a time tree. HAPPENS_AT is crucial for representing the time dimension in the graph model that is needed in time-varying graph queries and metrics for the analysis.

One important difference between the Geographical Space and Event Space is that numerical weights are only given to the edges of the Event Space. The edge weights are non-negative numbers that represent the time elapsed between two consecutive crash events. This is used to record the difference in time between successive traffic crash events which can used in the computation of weighted edge-metrics such as shortest paths between events.

The traffic accident influencing factors are also modeled as abstract entities and stored as separate nodes in the graph database. This makes easier for the traffic accidents to be classified in terms of influencing factors using graph queries and metrics. Each traffic crash can have one or more influencing factors which are connected the event space by the relationship "CAUSED_BY" edge.

### 3.4  Overview of the Graph Data Model

Fig. 1 provides a pictorial view of our proposed graph model. It depicts the meta-graph and conceptual representation of the entities in the model as implemented in the graph database

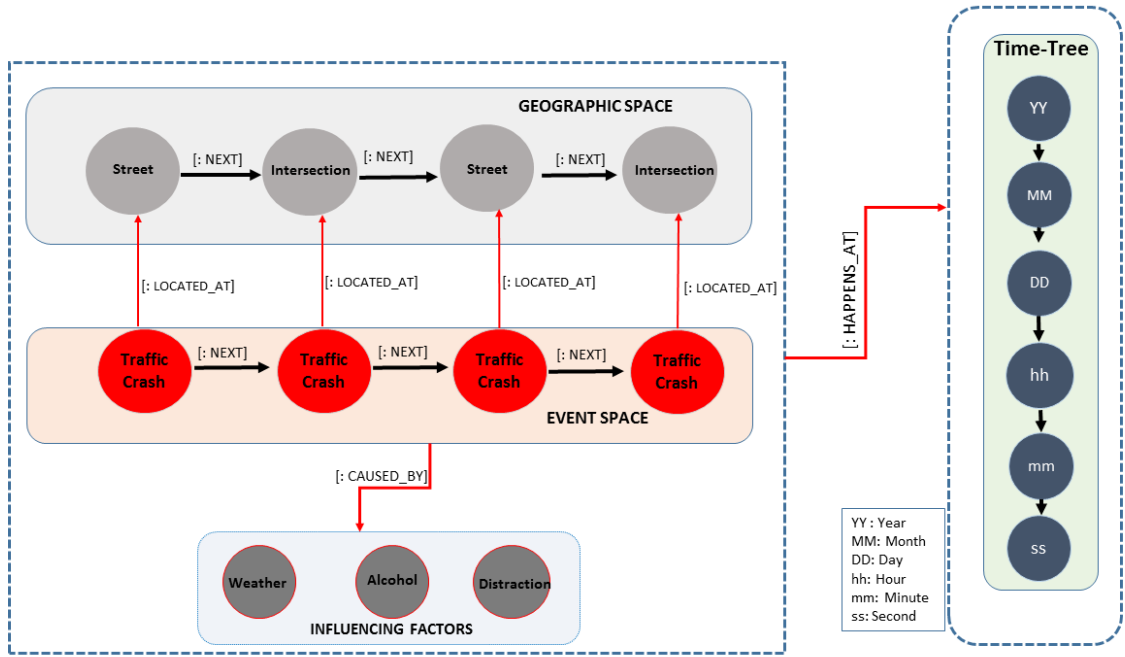

**Fig. 1** Overview of the proposed graph model

In the proposed model, the time dimension is not represented by a timestamp property associated with the nodes or edges of the Geographical and Event Spaces, but instead it becomes a relationship between the nodes in the network and the time tree. The time tree is a hierarchical and multi-level temporal indexing structure that represents the natural levels of a timestamp [44]. It is comprised of a "*Root*" node that links all the "*Year*" nodes, and in turn, each "*Year*" node is linked to the twelve "*Month*" nodes of a year, and so on. Fig. 2 illustrates this hierarchical structure where the top or lower-level nodes are connected through the "CONTAINS" relationship, meanwhile nodes situated at the same level of the time tree are sequentially connected through the "NEXT" relationship.

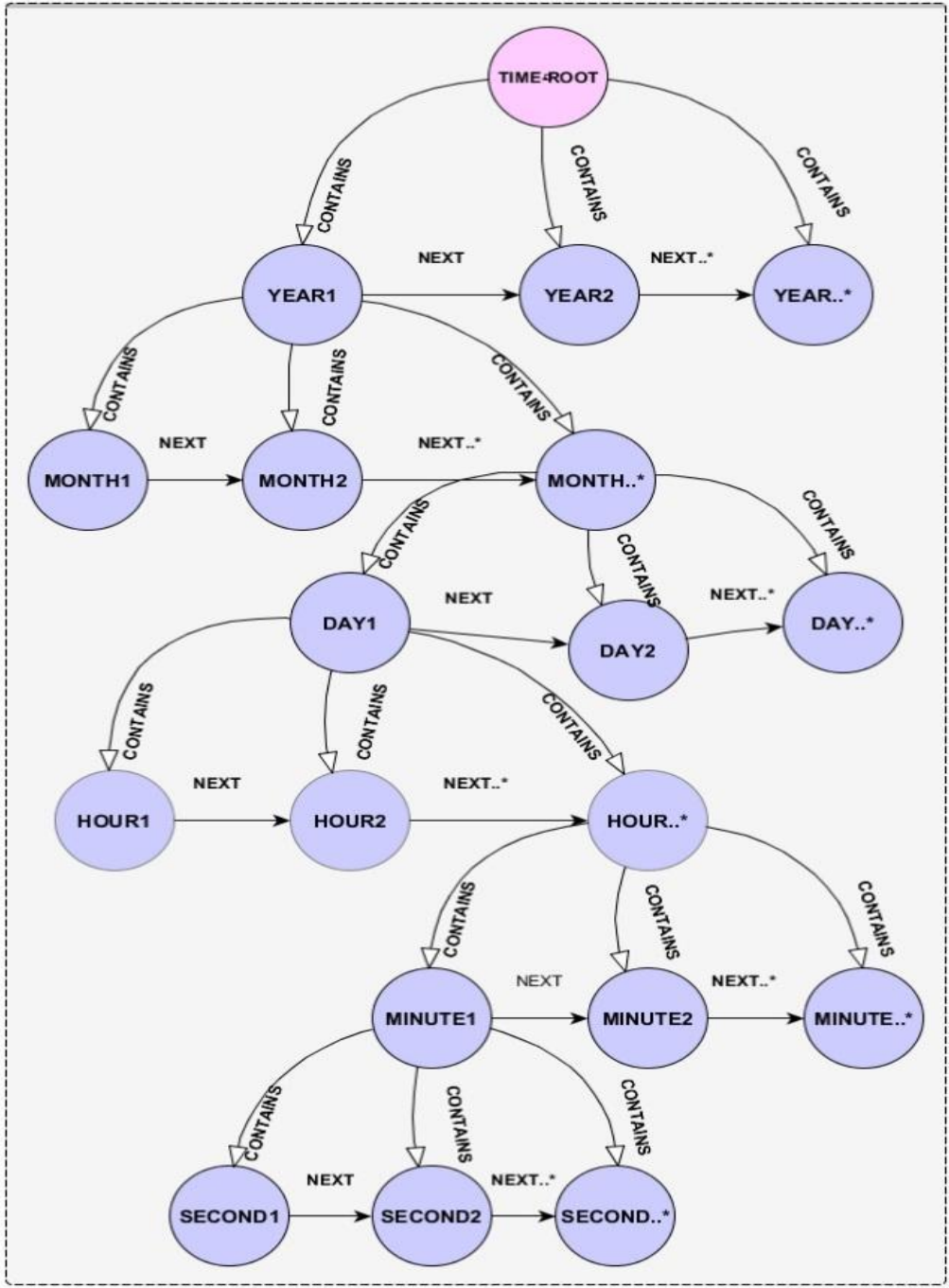

**Fig. 2** An example of the hierarchical indexing structure of the time tree [45]

This hierarchical temporal indexing structure which is a component of the model improves the processing speed of time-dependent graph queries. Essentially, a time-dependent query arrives at the time-instant node or the range of time-instant nodes of the query. Then it traverses to all the linked events or nodes without scanning through all the nodes in the large graph.

### 3.5 Graph Metrics for Analysis

There are many graph metrics such as connectedness, shortest and longest shortest paths, degree, betweenness and PageRank centralities as well as network diameter and density. However, to identify and rank traffic accident high-risk locations and accident activity pattern over time and space, degree and PageRank centrality metrics are the major focus in the study. These metrics are implemented and retrieved through incremental graph queries to the graph database. The incremental queries create projects subgraph footprints SF defined at time-points or time-intervals ($STVG_{t_1}, STVG_3, STVG_{t_3}, STVG_{t_4} \ldots \ldots STVG_{t_i}$).

#### 3.5.1 Degree centrality

Centrality measures give the relative measure of activity level a node has in the graph [46]. The Degree of a node is simply the amount of connectivity (the in-degree and out-degrees) that it has with the other nodes in the graph. The reputation of a node in the graph is enhanced if it has a high number of in-edges and out-edges. In this paper, degree centrality algorithm is utilized to identify traffic accident high-risk locations on streets and at intersections. Degree centrality is the simplest graph metric to compute but the most popular and extensively used in network sciences.

By definition: Degree centrality $C_D(i) = \sum_{j=1}^{n} a_{ij}$

Where, element $a_{ij} = 1$ (if a direct edge exists between nodes i and j) and $a_{ij} = 0$ (if there is no edge).

#### 3.5.2 PageRank centrality

PageRank was originally adopted by Google [47] for ranking of search results in the web network. PageRank is a variant of a more advanced view of centrality known as eigenvector centrality which permits connections in the network to have a variable value [48]. The results from PageRank centrality analysis most times correlates with that of degree centrality as both measure the level of importance of a node based on connectivity. However with PageRank, nodes are regarded as being more reputable if there are more incoming edges than outgoing edges, and the nodes that link to it are also important. In this paper, PageRank is used to rank accident high-risk locations over time and space.

The PageRank score of a node in STVG, n ∈ STVG was computed by aggregating the stationary probabilities of all the incident edges on the node.

PageRank of a node i, is defined as $PR(i) = d/n + (1 - d).\sum PR(j)/OutDegree(j)$ $(i \neq j \in N)$

Where d is the damping factor (ranging from 0 to 1), n is the total number of nodes and OutDegree is the number of outgoing edges from node j.

Incremental queries are used to implement these metrics at specific time-points and intervals on the projected subgraph footprints to evaluate traffic accident high-risk locations, activity level and ranking on the road network topology.

### 3.5.3 Dataset

The used case traffic accident dataset for this study was obtained from the free-access database of the Florida Department of Highway Safety and Motor Vehicles (FLHSMV) from 2010 to 2015-time coverage. The coverage area of this dataset is the Brevard County. Road network, boundary dataset and other ancillary spatial dataset were obtained from the Brevard County property appraiser free public dataset. The traffic accident dataset was downloaded as a shapefile that contained the 15 attributes as described in Table 1 and a total of 1048575 tuples. The tuples did not have any information about their spatial locations such as streets and intersections. Therefore we, had to carry out data-pre-processing and spatial contextualization.

**Table 1.** attributes of the traffic accident dataset

| Attribute | Description |
|---|---|
| Crash_DT | Crash date |
| Crash_TM | Crash time |
| Age | Age of the driver |
| Crash_HOD | Crash hour of the day |
| Crash_DOW | Crash day of the week |
| Crash_MOY | Crash month of the |
| Crash_Y | Crash year |
| Crash_WK | Crash week number of |
| Fatalities | Number of fatalities |
| Injury | Injury recorded |
| Alcohol_Related | Yes/No |
| Distraction_Related | Yes/No |
| Weather_Condition | Clear/Cloudy/Rain |

### 3.5.4 Data pre-processing and spatial contextualisation

Data pre-processing and spatial contextualization were vital to (i) add spatial context to the traffic accident datasets such as street name and intersect ID for seamless integration and connectivity at the database level. (ii) to decompose the intersections and streets in the entire Brevard County road network vector dataset into discrete vector unit, that is points (for intersection) and lixels (for street segments). These processes were carried out using some python script based on arcpy functions which involve an automated pipeline of geoprocessing as described in Fig. 3 and the following steps.

*Step 1: Computation of Intersections*: This involved a set of python functions coded in arcpy to; (i) merge streets and roads of the same unique name that are disconnected in the road network dataset into single streets and roads. (ii) a spatial search function is used to search for points where two or more streets or roads are intersected. (iii) these intersected points are automatically extracted into a list with unique IDSS numbers and a concatenation of the street names that formed the intersection. (iv) duplicate points are filtered, removed and the final list of points saved as a vector layer called intersections.

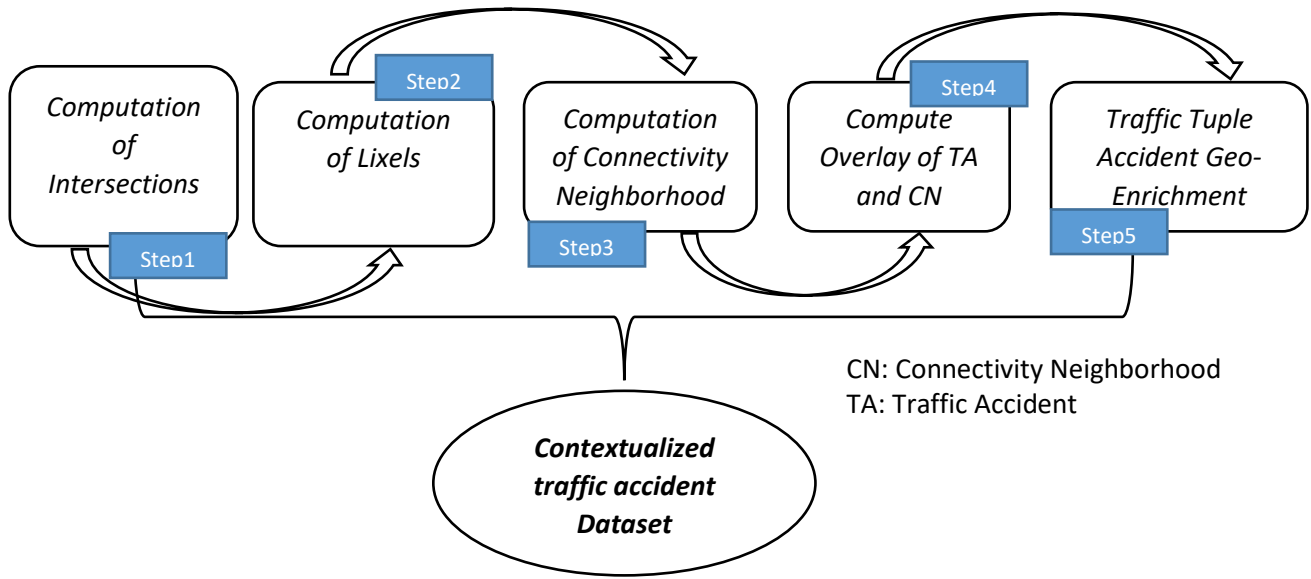

**Fig. 3** Data pre-processing and contextualization workflow

*Step 2: Computation of Lixels:* Each street or road segment is divided into a basic linear unit of a 50 meters length. These linear units of segments are called Lixel as defined by Xie and Yan [39]. This is to facilitate a systematic selection of regularly spaced locations along the street network to enhance connectivity to discrete street nodes in the graph model.

*Step 3: Connectivity Neighborhood Creation:* This step creates a 15-m connectivity radius around each street lixel and intersection point. This is used to associate the traffic accident events to their respective streets and intersections within 15 meters radius of the accident epicenter. These connectivity neighborhoods are converted to a vector feature layer and each search neighborhood is labeled by the name of its associated streets lixel of intersection.

*Step 4. Overlay Traffic Accident and Connectivity Neighborhood:* A spatial overlay function is called up from arcpy python package to associate each traffic accident tuple to its respective connectivity neighborhood.

*Step 5: Geo-enrich the Traffic Accident tuple:* Each traffic accident tuple is geo-enriched by adding as extra attribute of this tuple, the label of the associated connectivity neighborhood which is either a street name or an intersection ID. Finally, the spatially contextualized traffic accident dataset is realized to fit the graph model described section 3.4.

*Step 6: Temporal Sequence of Occurrence:* The last preprocessing carried out on the contextualized traffic accident dataset is sequential arrangement of the events based on temporal sequence of occurrence. An additional column was added to the CSV file of the dataset called "Sequence". This enables us to create the "NEXT" space-time relation both in the event space and geographic space in the graph database.

### 3.5.5 Neo4j graph database

The implementation of our graph model was done using the Neo4j graph database management system. Neo4j is a popular native graph database with applications in fraud detection, real-time recommendation engines, and network operations. Neo4j uses the Cypher query language which enables ad-hoc queries and user defined functions (UDFs). One important characteristic of the Neo4j database is to create the edges of the network at the same time as the nodes are being created, rather than creating them at query time through JOIN operations. Table 2 shows an example node creation cypher statements to load the traffic accident tuples into Neo4j database.

**Table 2** Cypher code used to create Traffic accident nodes

*load csv Traffic Crashes as cr*
*CREATE (:Crashes {CrashID: cr.FID, CrashTime: cr.CRASH_TM,*
*CrashAge: cr.AGE, Hour:toInt(cr.CRASH_HOD),*
*WDay: cr.CRASH_DOW, Month: toInt(cr.CRASH_MOY) ,*

*Year:toInt(cr.CRASH_YEAR),Sequence=toInteger(cr.sequence),Fatalities: cr.Fatalities,Injuries: cr.Injuries, Alcohol_Re: cr.Alcohol_Re, Distraction_Re: cr.Distractio, Weather_Condition: cr.Weather_Co, StreetName:*
*cr.ST_NAME, IntName: cr.INT_ID, Day: toInt(substring(cr.CRASH_DT,0,2))});*

After the nodes have been created, the edges can be created in the database. Table 3 shows an example of the edges created between the nodes of the Geographical Space and Event Space. A total of 1137695 nodes and 344765 edges were created in few seconds in the Neo4j database.

**Table 3** Cypher code for creating the link between the Geographic and Event Spaces.

*///connect crashes and streets*

*MATCH (cr:Crashes), (st:Streets) where cr.StreetName = st.StreetName*

*MERGE (cr)-[: LOCATED_AT]->(st);*

*///connect crashes and intersection*

*MATCH (cr:Crashes), (int:Intersections) where cr.IntID = int.IntID*

*MERGE (cr)-[: LOCATED_AT]->(int);*

The nodes were sequentially linked to the time tree from the highest to the lowest level of time granularity as shown in Table 4. This Cypher code was used to create the sequential "HAPPEN_AT" edge between the Traffic accident nodes and the corresponding time tree leaf nodes. In total, 109844 edges have been created for "HAPPEN_AT" edges within the database.

**Table 4** Step-wise linking of crash nodes to the Time-Tree.

*MATCH (cr:Crashes) WITH cr*

*MATCH (yy:Year {yearid:cr. Year}) WITH cr,yy*

*MATCH (yy)-[r1]->(mm:Month {monthid:cr.Month}) WITH cr,yy,mm*

*MATCH (mm)-[r2]->(dd:Day {dayid:cr.Day}) WITH cr,yy,mm,dd*

*MATCH (dd)-[r3]->(hh:Hour {hourid:cr.Hour}) WITH cr,yy,mm,dd, hh*

*CREATE (cr)-[:HAPPENED_AT]->(hh);*

### 3.5.6 Cypher queries

Cypher is a graph query language that allows for efficient querying, updating and analysis of graph properties and metrics. Graph metrics (e.g., PageRank and degree centralities), and other UDFs are easily encoded within cypher queries. The computation of the graph metrics used in this study was coded in cypher with a pipeline of query functions. Table 5 illustrates this pipeline for a PageRank cypher query. Through the Cypher query manager, cypher queries are used to project the part of the graph valid for the specified time-window in-memory. Appropriate graph metrics is then implemented on the projected graph to retrieve analytical values needed traffic accident high-risk location analysis.

**Table 5** Time-incremental PageRank query.

---

***Step 1: Retrieve only the nodes and edges from 1am to 12pm***

*WITH range (1,24) AS Hour*

***Step 2: Compute the PageRank for every hour incrementally***

*FOREACH (hour IN Hour |*

*CALL PageRank (({*

***Step 3: Write the values to the nodes as their property 'name'***

*Write: True,*

*Property: 'PageRank',*

***Step 4: Find the "accident" nodes which are located at an Intersection***

*{Node:'MATCH (int:Intersection) RETURN id(int)AS id',*

*Relationship:'MATCH (d: Day)- [: HAS_HOUR] -(h:Hour)<-[r1: HAPPENS_AT]-(cr:Crash) WITH cr*

*MATCH (cr)-[r]-(int: Intersection)}*

***Step 5: Return the values of PageRank as the following node definition***

*RETURN id(cr) AS source, id(int) AS target, count (*) AS PageRank ORDER BY PageRank'}))*

---

Table 6 Time-incremental Degree-Centrality query.

*WITH range (1,24) AS Hour*

*'MATCH (d: Day)- [: HAS_HOUR] -(h:Hour)<-[r1: HAPPENS_AT]-(cr:Crash) WITH cr*

*MATCH p = (cr:Crash)-[r:NEXT*..]-(cr:Crash)*

*UNWIND NODES (p) [1..-1] as n*

*WITH n, NODES (p) as nodes FILTER (node in nodes*

*WHERE (node:Intersection)) as int RETURN*

*DISTINCT(int), size((int)-[]-(n)) as degree*

## 4. Discussion of Results

Identification, ranking and profiling of traffic accident high-risk locations have always been the major concerns of most transport authorities and traffic accident departments. They also want to be able to carry out this analysis at varying spatial and temporal scales efficiently with simple queries with less complexities. Current analysis methods based on GIS and spatial statistics are not only modeled without temporal context in them but can be very challenging and technically complex for a traffic agent to process and make sense of the analytical results. Graph metrics such as degree and PageRank can be implemented in simple graph queries to the database with temporal and spatial context in the queries and results retrieve faster and easy to understand than in the case of any GIS and Statistical analytics tools.

The analysis and results below, present the use of degree and PageRank graph metric queries such as shown in Table 7 to answer the following questions: Which intersections and streets have the highest crash index? When and where do most crashes occur? how does spatio-temporal pattern of fatalities differ from spatio-temporal pattern of traffic accidents overall? What is the spatio-temporal pattern during weekdays and weekends, morning, afternoon and night times and so on.

**Table 7** Example Queries to compute and rank average crash index of intersections

| MATCH (cr: Crashes)-[r]-(int:Intersections) RETURN int.IntersectionID, int.IntersectionName, Size((int)-[]-() ) AS degree ORDER BY degree DESC LIMIT 20; | MATCH (int:Intersections) WITH collect(int) AS nodes CALL apoc.algo.pageRank(nodes) YIELD node, score RETURN node. IntersectionID, node.IntersectionName, score ORDER BY score DESC LIMIT 20; |
|---|---|

**Table 8** Top-20 intersections with the highest crash activity based on degree centrality graph metric

| Intersection Name | Intersection ID | Degree Centrality |
|---|---|---|
| Ramp&Ramp&Ramp | 1886 | 3565 |
| I 95&I 95 | 12639 | 3355 |
| Highway 1&Turnaround&Highway 1 | 1817 | 1585 |
| Ramp&I 95&I 95 | 14472 | 1099 |
| Highway 1&Highway 1&Turnaround | 12764 | 763 |
| Rockledge&Rockledge&Turnaround | 23264 | 659 |
| Highway A1A&Highway A1A | 20243 | 561 |
| Wickham&Wickham | 25863 | 547 |
| Ramp&Ramp | 3485 | 531 |
| Washington&Turnaround&Washington | 13047 | 469 |
| Minton&Palm Bay&Minton&Palm Bay | 25495 | 391 |
| Wickham&Sarno&Sarno&Wickham | 22107 | 301 |
| Eau Gallie&Eau Gallie&Harbor City&Harbor City | 23012 | 276 |
| Minton&Emerson&Emerson&Minton | 23851 | 269 |
| Turnaround&Minton&Minton | 13591 | 256 |
| Eau Gallie&Eau Gallie | 1821 | 249 |
| Atlantic&Atlantic&Fourth&Fourth | 22580 | 247 |
| Eau Gallie&Eau Gallie&South Patrick&Riverside | 24088 | 218 |
| Turnaround&Courtenay&Courtenay | 14833 | 216 |

**Table 9** Top-20 street-segments with the highest crash activity and Ranking based on degree centrality and PageRank graph metrics respectively

| Street Name | PageRank Score | Degree Centrality |
| --- | --- | --- |
| I 95 | 526.27865 | 7018 |
| Babcock | 462.7941 | 6203 |
| Wickham | 313.8318 | 4181 |
| Highway 1 | 279.09709 | 3805 |
| Atlantic | 255.17327 | 3811 |
| Harbor City | 241.52099 | 3380 |
| Malabar | 219.22723 | 3036 |
| Eau Gallie | 210.35931 | 3099 |
| New Haven | 208.55399 | 2716 |
| Courtenay | 201.5698 | 2882 |
| Palm Bay | 172.51817 | 2297 |
| Merritt Island | 129.41375 | 1773 |
| Highway A1A | 121.02052 | 1748 |
| Cheney | 102.26838 | 1299 |
| Emerson | 102.19247 | 1583 |
| Minton | 94.63176 | 1373 |
| Ramp | 93.52406 | 1248 |
| Cocoa | 89.97375 | 1222 |
| Fiske | 80.78942 | 1231 |
| Dairy | 80.7024 | 1152 |

In this analysis, degree centrality graph measure is very efficient in identifying traffic accident high-risk locations both on the road network segments (streets) and on intersections only based on network connectivity between the geographic space and the event space. Fig. 4 shows the graph visualization of these interactions. Our analysis identified that intersections with ID 1886 and 12639 on the Ramp and I 95 roads (Fig. 5) with degree of 3565 and 3355 respectively shown in Table 10 have the highest traffic crash activity over the 6-year period.

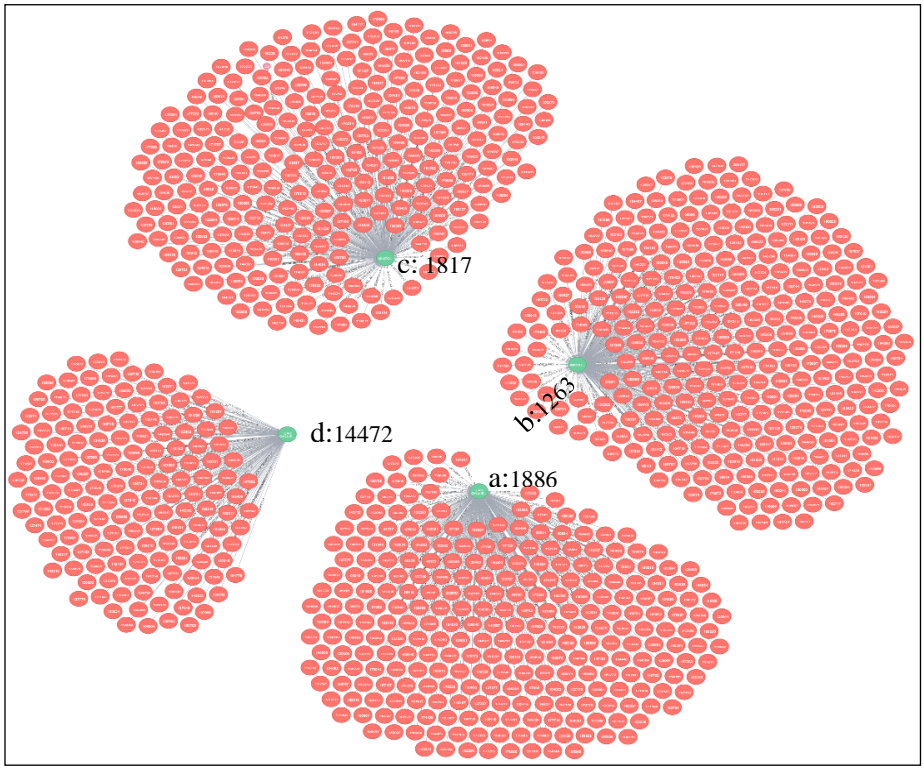

**Fig. 4** Graph visualization of the connectivity between road-network intersections and traffic crashes. Intersections are the green nodes and the red nodes are the crashes

I 95 and Babcock roads (Fig. 4) are the most traffic accident-prone road segments with degree values of 7018 and 6203 respectively within a period of 6 years.

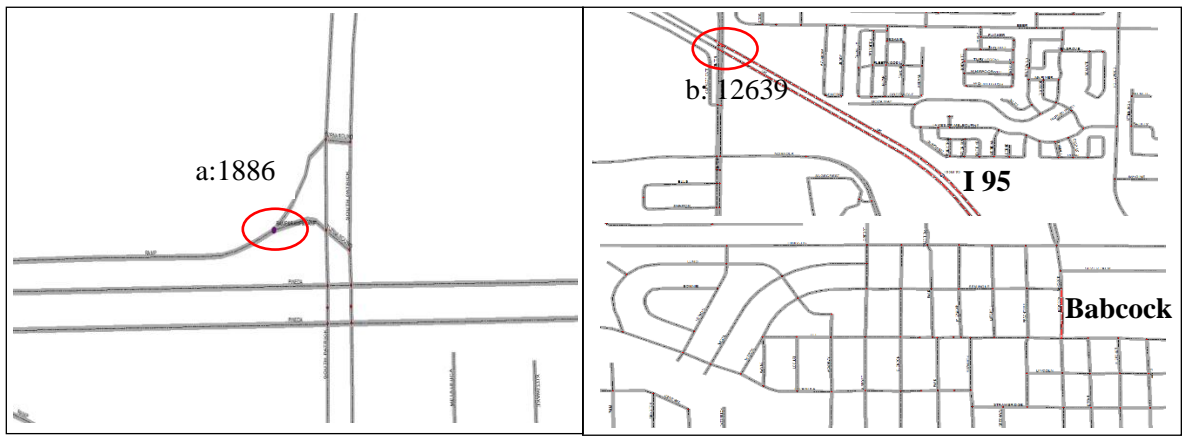

**Fig. 5** Some streets and intersections shown on the road network topology; I 95 and Babcock streets in red lines while the intersections are circled in red

However, as the degree of centrality employs in-edges and out-edges for its computation, PageRank gives more relevance to the in-coming edges and the importance of the nodes that link to the node in question. PageRank values as shown in Table 9 discovers similar results in terms of traffic accident streets ranking over a period of 6 years. However, the top 20 intersections with the highest degree centrality measures are not ranked top in the PageRank analysis shown in Table 10. This is because PageRank does not only consider the connectivity between the intersections and the traffic crash events but also factors in the degree of importance of that intersection in the entire network. That is the number of roads or street segments that have a direct connectivity with any intersection in question. The highly ranked intersections have more segment connections than others. For example, the top 10 ranked spots in this analysis have at least 4 connecting segments. PageRank metric also considers some intersections very important because they are high-risk locations and have direct connections to high-risk street segments in the network.

**Table 10** Top-20 intersections with the highest crash activity ranking based on PageRank graph metric

| Intersection Name | Intersection ID | PageRank Score |
|---|---|---|
| Wickham&Sarno&Sarno&Wickham | 13047 | 14.06875 |
| Eau Gallie&Eau Gallie&Harbor City&Harbor City | 23012 | 13.00625 |
| Minton&Emerson&Emerson&Minton | 23851 | 12.5175 |
| Eau Gallie&Eau Gallie&South Patrick&Riverside | 24088 | 10.35 |
| Harbor City&Harbor City&Strawbridge&Strawbridge | 24586 | 9.90375 |
| Interchange&San Filippo&Malabar&Malabar | 22399 | 9.33 |
| Aurora&Wickham&Wickham&Aurora | 23190 | 8.92625 |
| Wickham&Lake Washington&Wickham&Lake Washington | 21677 | 7.77875 |
| Barnes&Murrell&Murrell&Murrell&Barnes | 25766 | 7.5875 |
| Minton&Palm Bay&Minton&Palm Bay | 25495 | 6.81111 |
| Minton&Palm Bay&Minton&Palm Bay | 25523 | 6.81111 |
| Hibiscus&Hibiscus&Babcock&Babcock&Babcock | 25758 | 6.71625 |
| Harbor City&Sarno&Harbor City&Harbor City | 22971 | 6.67375 |
| Garden&A Max Brewer Memorial&Washington&Washington | 22938 | 6.27 |
| Merritt Island&Newfound Harbor&Merritt Island&Merritt Island | 25736 | 5.675 |
| Wickham&Parkway&Wickham&Wickham&Parkway | 25761 | 5.675 |
| Cypress&Eau Gallie&Cypress&Montreal&Eau Gallie | 23255 | 5.6325 |
| St Lucie&Atlantic&Atlantic&Atlantic&St Lucie | 25863 | 5.56875 |
| Atlantic&Atlantic&Turnaround&Marion | 25371 | 5.52625 |
| Otterbein&Clearlake&Clearlake | 6745 | 5.48375 |

It is shown in the Fig. 6, the number of accidents that take place yearly on these roads especially I 95 and Babcock Street maintained a record high throughout these 6 years as well as in every month of the year. However, the analysis shows that a street as small as Babcock Street has maintained the highest degree of traffic accidents since 2013 consistently. Based on these metrics it becomes easy for a traffic accident agent to recommend Babcock Street for traffic accident profiling as it has maintained the highest degree of accidents since 2013 on yearly and monthly basis. The intersection with ID, 1886 on Ramp Road has maintained the top-accident prone intersection over the 6 years, except in 2014, that the intersection with ID, 12639 on I 95 road hit a record high over all other intersections in the network. This suggest that the spot was probably in a bad shape in 2014. Coincidentally, no accident has been recorded on this intersection, 12639 in August and September throughout these 6 years. Interestingly, the months of October, November and December are relatively the months with the lowest degree of traffic crashes both on the road segments and intersections.

Fig. 7 shows daily analysis which depicts consistent peaks on Mondays and Fridays for both on the segments and intersections. However, the two most high-risk accident intersections 1886 and 12639 maintained average highest on a daily similar to the Babcock and I 95 roads. These traffic accident peaks are recorded mainly during peak hours of the day as shown in the hourly charts. The highest peaks for the crashes are recorded between the hours of 3 pm and 7 pm daily. Time-varying PageRank analysis did not show significant difference from the overall PageRank average described in Tables 8 and 9.

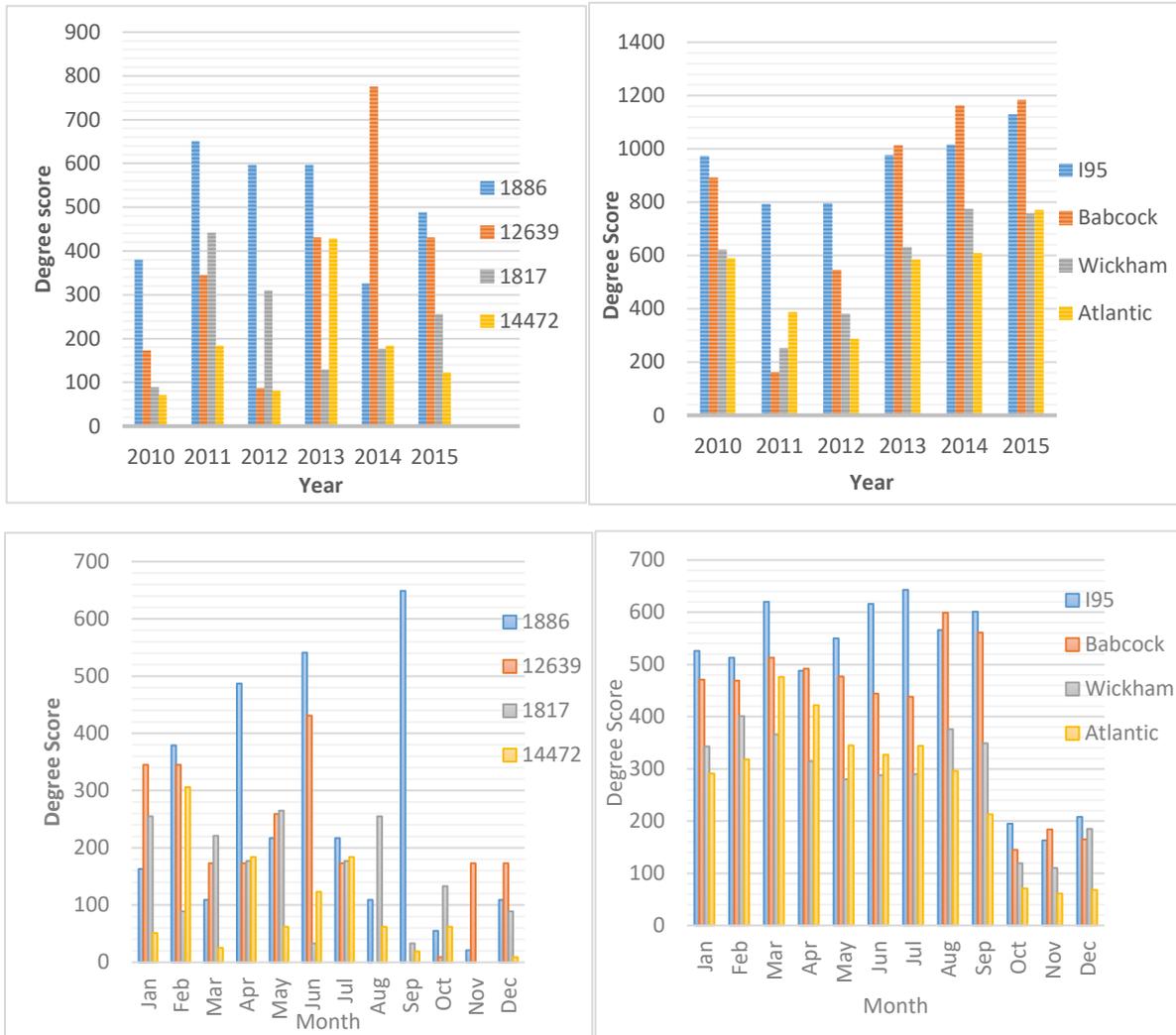

**Fig. 6** Top four streets and intersections traffic accident degree scores varying yearly (top) and monthly (bottom). Intersections on the right and Streets on the left

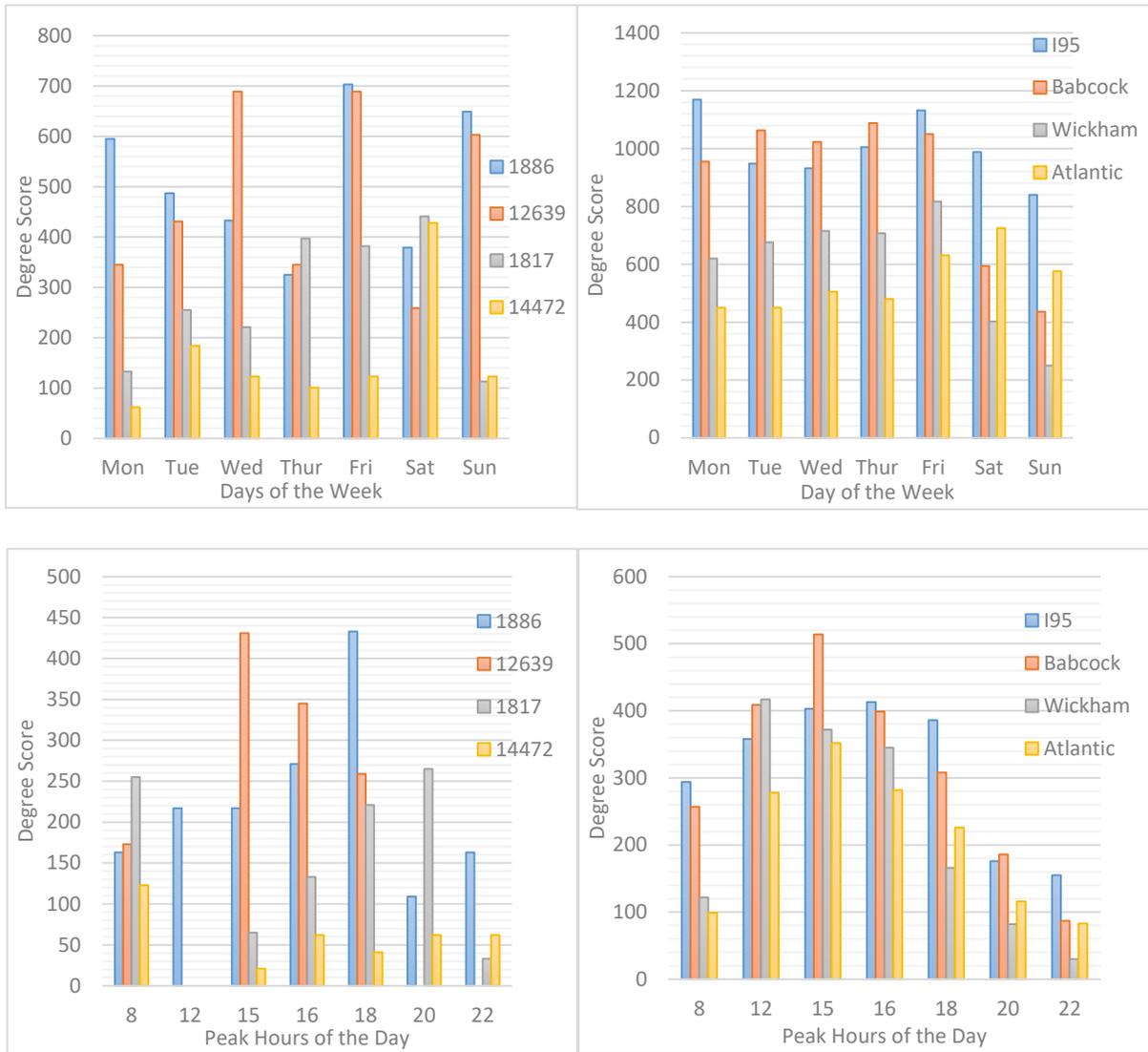

**Fig. 7** Top four streets and intersections traffic accident degree scores varying daily (top) and hourly (bottom). Intersections on the right and Streets on the left

Fig. 8 shows that most accidents in intersection 1886 were not fatal as it did not appear in the top 20 high traffic accident fatality charts. However, the intersection, 13462 on Highway 1 which ranks third in degree of accident occurrence has recorded most fatality in Brevard County from 2010-2015. I95 road maintained the highest in terms of degree of accidents, ranking and fatality rate.

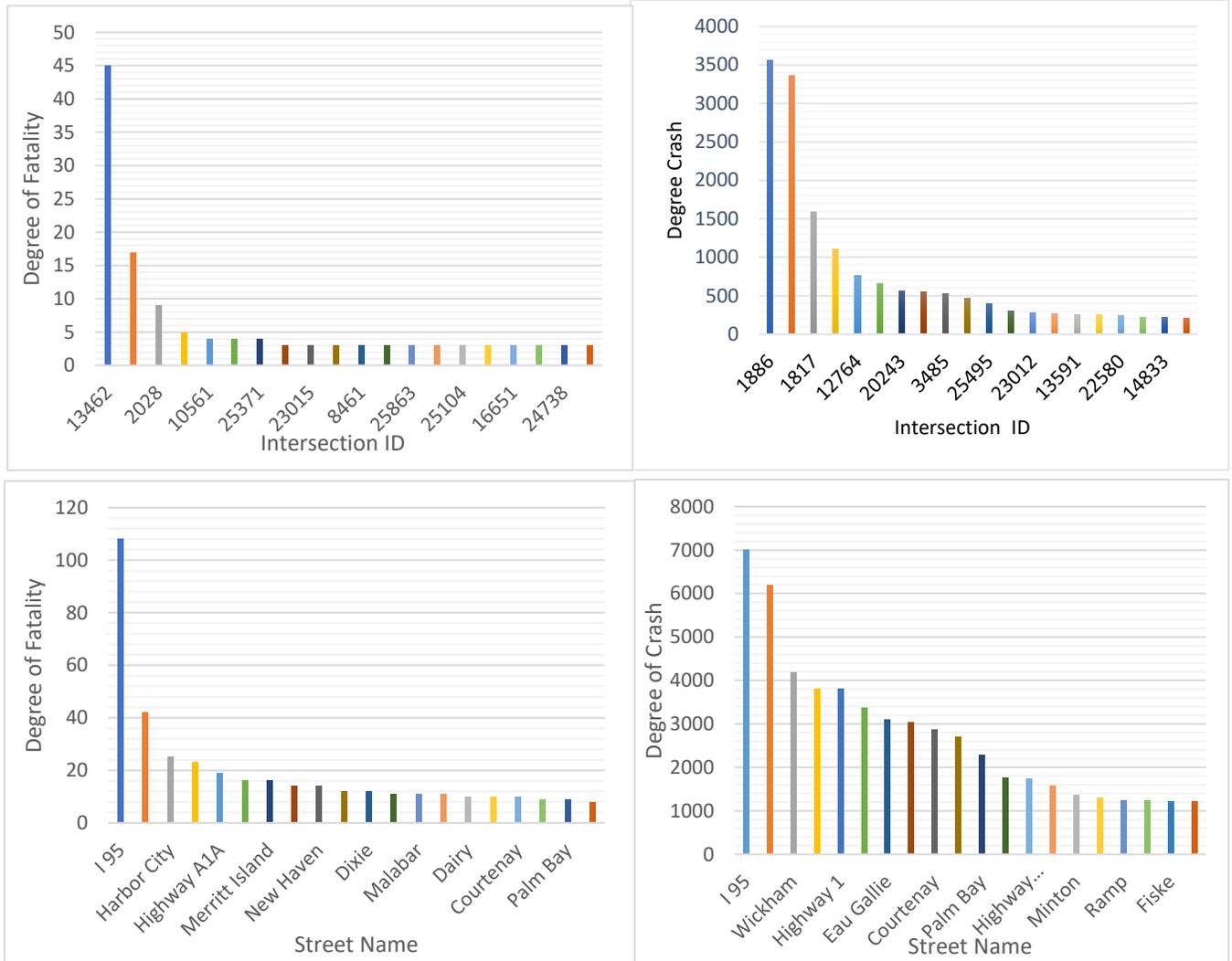

**Fig. 8** Average traffic accident Degree of fatality vis-a-vis overall Degree of crashes on the intersections and streets in Brevard County over a period of 6 years

## 5. Conclusion

While previous research on spatio-temporal traffic accident pattern analysis have mostly been based on GIS-based methods and spatial point pattern analysis, in this paper, we have come from the perspective of space-time interactions in the entire network space of road network topology and traffic accident events. In other words, this study has focused on discovering traffic accident patterns in space and time based on graph analytics of the network that emerges from the space-time-varying relationship between road network and traffic accident events.

This paper describes a simple space time-varying graph (STVG) model of traffic accident that is generic and scalable. The STVG model provides us with the capability to identify, rank and profile traffic accident high-risk locations in space and study their pattern over time. The developed STVG model was implemented in the Neo4j graph database with inbuilt graph algorithms. The time-tree is hierarchical temporal indexing structures that provide the ability to analysis these patterns at different space and time scales. Data spatial contextualization became very necessary so that the traffic accident dataset could fit in properly in the space time-varying graph model. And this was carried out through a pipeline of spatial data preprocessing as described in section 3.5.4.

Traffic accident blackspots and segments were identified and ranked in space and time using time-dependent degree and PageRank centrality graph metrics. Degree centrality is simple and effective in discovering these high-risk locations and segments in space and time while PageRank metric ranking adopted a more complex procedure as explained in the section 3.52. The PageRank of the street segments correlates with the degree centrality analysis but does not correlate with degree centrality analysis of the intersections. PageRank metric requires some modification to suit this type of network, however, degree centrality metric proved to be most suitable.

Further analysis will incorporate more of the non-spatial and temporal entities of the network such as the influencing traffic accident factors and fatality in the analysis. Analysis that answers question such as when and where are the blackspots for crashes involving elderly drivers, teenagers, alcohol or weather related? can be answered based on this model.

**Conflict of Interest**

On behalf of all authors, the corresponding author states that there is no conflict of interest.